\documentclass{llncs}
\usepackage{gnuplot-lua-tikz}
\usepackage{url}  
\def\miko{Mikol\'a\v{s} Janota}
\def\theTitle{QFUN: Towards Machine Learning in QBF}
\def\theTitleBr{QFUN: Towards Machine Learning in QBF}
\def\theAuthor{\miko}
\title{\theTitleBr}
\author{\theAuthor}
\institute{IST/INESC-ID, Portgual\\\email{mikolas.janota@gmail.com}}
\usepackage{xspace}
\def\rareqs{\textrm{RAReQS}\xspace}
\def\areqs{\textrm{AReQS}\xspace}
\def\qfun{\textrm{qfun}\xspace}
\def\qfunT{\textrm{qfun}\ensuremath{^2}\xspace}
\usepackage{amsmath,amssymb}
\usepackage{stmaryrd}
\newtheorem{observation}{Observation}
\newcommand{\comprehension}[2]{\ensuremath{\left\{ {#1} \mid {#2}\right\}}}
\newcommand*{\ite}[3]{{#1}\,?\,{#2}\,:\,{#3}}
\newcommand{\assgn}[2]{{#2{\shortrightarrow}#1}}

\DeclareMathOperator*{\Q}{\mathcal{Q}}

\DeclareMathOperator*{\lev}{{\sf lv}}
\DeclareMathOperator*{\var}{{\sf var}}
\DeclareMathOperator*{\vars}{{\sf vars}}
\DeclareMathOperator*{\union}{\cup}
\DeclareMathOperator*{\dom}{{\sf dom}}

\DeclareMathOperator*{\Is}{\mathcal{I}}
\DeclareMathOperator*{\samples}{\mathcal{E}}
\usepackage[linesnumbered,ruled,vlined]{algorithm2e} 
\let\oldnl\nl
\newcommand{\nonl}{\renewcommand{\nl}{\let\nl\oldnl}}
\SetKwFunction{SAT}{SAT}
\SetKwFunction{solve}{Solve}
\SetKwFunction{refine}{Refine}
\SetKwFunction{Break}{Break}
\SetKwFunction{Continue}{Continue}
\SetKwFunction{learn}{Learn}
\SetKwFunction{magic}{ShouldLearn}
\SetKwFunction{refine}{Refine}
\SetKwFunction{encode}{EncodeNeg}
\SetKwFunction{clauseMap}{clauseMap}
\SetKwFunction{prenex}{Prenex}
\SetKwFunction{test}{Test}
\SetKwData{out}{outc}%
\SetKwData{nul}{$\bot$}%
\SetKwData{sfalse}{false}
\SetKwData{strue}{true}
\SetKwData{hasCEX}{hasCM}
\SetKwInOut{Input}{input}
\SetKwInOut{Output}{output}
\DontPrintSemicolon%
\SetKw{Let}{let}
\SetKw{Func}{Function}
\SetKwFunction{addvariable}{addVars}
\SetKwFunction{addformula}{addFla}
\SetKwData{out}{outc}%
\SetKwFunction{winsOne}{Wins1}
\usepackage{color}
\definecolor{citeblue}{rgb}{0.1,0,.4}
\definecolor{refcolor}{rgb}{0,0,0.4}
\definecolor{midgreen}{RGB}{0,150,0}
\definecolor{darkgreen}{RGB}{0,128,0}
\definecolor{darkblue}{RGB}{0,0,128}
\definecolor{darkred}{RGB}{192,0,0}
\usepackage[pdftex%
,colorlinks=true%
,bookmarks=false%
,linkcolor=citeblue%
,citecolor=citeblue%
,urlcolor=blue%
,plainpages=false]{hyperref}
\hypersetup{%
pdfauthor={{\miko}},
pdftitle={{\theTitle}}
}

\begin{document}
\maketitle
\begin {abstract}
This paper reports on the QBF solver QFUN that has won the non-CNF
track in the recent QBF evaluation.
The solver is motivated by the fact that it is easy to construct Quantified
Boolean Formulas (QBFs) with short winning strategies (Skolem/Herbrand
functions) but are hard to solve by nowadays solvers.  This paper argues that a
solver benefits from generalizing a set of individual wins into a strategy.
This idea is realized on top of the competitive RAReQS algorithm by utilizing
machine learning.  The results of the implemented prototype are highly
encouraging.
\end {abstract}

\section{Introduction}
   Against all odds posed by computational complexity,
   logic-based problem solving had a remarkable success at research but also industrial level.
   One of the impressive success stories is the Boolean  satisfiability problem (SAT).
   Quantified Boolean formulas (QBF) go one step further and extend  SAT with quantification.
   This enables targeting a larger class of problems~\cite{BenedettiJSAT08,RintanenAAAI07,Umans-sigact02,Giunchiglia-faia09}.
   However, success of QBF solvers comparable to SAT still seems quite far.
   Nevertheless, we have recently seen a significant progress in the area almost every year,
   e.g.
   ~\cite{zhang-iccad02,Biere04,Benedetti05,TuHJ15,GoultiaevaB10,Goultiaeva-date13,JanotaIJCAI15,RabeFMCAD15,JanotaKMC12,LonsingSAT16,TentrupSAT16,RabeSAT16}.
   This paper aims to make a case for the use of machine learning \emph{during} QBF solving.

   It has been observed that search is often insufficient.
   A well-known example is the formula
       $\forall X\exists Y.\,\bigwedge x_i\leftrightarrow y_i$ with
       $X=\{x_1,\dots,x_n\}$ and $Y=\{y_1,\dots,y_n\}$~\cite{letz-tableaux02}.
%
   Traditional search will  easily find an assignment (valuation) to $X$ and $Y$ satisfying the matrix (the propositional part).
   However, to prove that there is an assignment for $Y$ given \emph{any} assignment to $X$ is difficult.
   Traditional search, even with various extensions, will try exponentially many assignments.
   A human can easily see why the formula is true.
   Indeed, given an arbitrary assignment to $X$, setting each $y_i$ to $x_i$ gives a witness for the validity of the formula.

   It is useful to see QBFs as two-player games, where the
   existential player tries to make the formula true and the universal
   false. A winning strategy for the existential player
   shows that it is true.  The formula above is a good example of a \emph{small
   winning strategy}---the strategy for~$y_i$ is the function $s_{y_i}(x_1,\dots,x_n)\triangleq x_i$.
   The million dollar question here is, where do we get the strategies?
   This paper builds on the following idea:
   {\it Observe a set of assignments and learn from them strategies using machine learning.
   In another words, rather than looking at individual assignments, collect a set of them and generalize them into a strategy.}

   Learning a strategy is not enough---it must also be
   incorporated into a solving algorithm.  A straightforward approach would
   be to test for the learned strategy whether it is a winning
   one (that is possible with a SAT call~\cite{BuningSZ07}). However, this
   would put a lot of strain on the learning since we would have to be quite
   lucky to learn the right strategy and eventually we would have to deal with large training sets.

   The algorithm presented in this paper takes  inspiration in the existing  algorithm \rareqs,
   which gradually expands the given formula by plugging in the encountered assignments~\cite{JanotaKMC16}.
   Instead of plugging in assignments, we will be plugging in the learned strategies.
   This forms the second main idea of the paper:
       {\it  Expand the formula using strategies learned from collected samples
       and then start collecting a new set of samples.}


\section{Preliminaries}\label{sec:preliminaries}

A \emph{literal} is a Boolean variable or its negation. The literal
complementary to a literal $l$ is denoted as $\bar l$, i.e.\ $\bar
x=\lnot x$, $\overline{\lnot x}=x$.
For a literal $l=x$ or $l=\lnot x$, we write $\var(l)$ for $x$.
Analogously, $\vars(\phi)$ is the set of all variables in formula $\phi$.
An assignment is a mapping from variables to Boolean constants $0,1$.
For a formula $\phi$ and an assignment $\tau$, we write $\phi[\tau]$
for the  substitution of the variables in the domain of $\tau$ with their respective  constants.

\subsection{Quantified Boolean Formulas}

\emph{Quantified Boolean Formulas}
(QBFs)~\cite{DBLP:series/faia/BuningB09} extend propositional logic by
enabling quantification over Boolean variables. Any propositional
formula $\phi$ is also a QBF with all variables \emph{free}. If $\Phi$
is a QBF with a free variable~$x$, the formulas $\exists x.\,\Phi$ and
$\forall x.\,\Phi$ are QBFs with~$x$ \emph{bound}, i.e.\ not free. Note
that we disallow expressions such as $\exists x.\exists x.\,x$, i.e., each variable is bound at most once. Whenever
possible, we write $\exists x_1\dots x_k$ instead of $\exists
x_1\dots\exists x_k$; analogously for~$\forall$. For a QBF
$\Phi=\forall x.\,\Psi$ we say that $x$ is \emph{universal} in~$\Phi$
and is existential in $\exists x.\,\Psi$. Analogously, a literal $l$
is universal (resp.\ existential) if $\var(l)$ is universal
(resp.\ existential).

Assignments also can be  applied to QBF
with
$(Qx.\,\Phi)[\tau]$ defined as $\Phi[\tau]$
if $x$ is in the domain of $\tau$ with $Q\in\{\forall,\exists\}$.

A QBF correspond to a propositional formula: $\forall x.\,\Psi$ corresponds to
$\Psi[\assgn{0}{x}]\land\Psi[\assgn{1}{x}]$ and $\exists x.\,\Psi$ to
$\Psi[\assgn{0}{x}]\lor\Psi[\assgn{1}{x}]$.  Since $\forall x\forall
y.\,\Phi$ and $\forall y\forall x.\,\Phi$ are semantically equivalent,
we allow $Q X$ for a set of variables $X$, $Q\in\{\forall,\exists\}$.
A QBF with no free variables is \emph{false}
(resp.\ \emph{true}), iff it is semantically equivalent to the constant
$0$ (resp.\ $1$).

A QBF is \emph{closed} if it does not contain any free variables.  A
QBF is in \emph{prenex form} if it is of the form
\hbox{${\Q}_1 X_1 \dots {\Q}_k X_k.\,\phi$},
where ${\Q}_i\in\{\exists,\forall\}$, ${\Q}_i\neq {\Q}_{i+1}$,
$\phi$ propositional, and $X_i$ pairwise disjoint sets of variables.
The propositional part~$\phi$ is called the
\emph{matrix} and the rest \emph{prefix}.  For a variable $x\in X_i$
we say that $x$ is at \emph{level} $i$ and write
$\lev(x)=i$; we write $\lev(l)$ for $\lev(\var(l))$.
Unless specified otherwise, QBFs are assumed to be closed and in prenex form.

\subsection{Games and Strategies}\label{sec:games}
For most of the paper, QBFs are seen as two-player games.
The existential player tries to make the matrix true
and conversely the universal player to make it  false.
A player assigns value only to variables that belong to the player and may
assign a variable only once all variables that precede it in the prefix are
assigned. In other words, the two players assign values following the
order of the prefix.
The game semantic perspective on QBF has the advantage that mostly
we do not need to distinguish between the player $\exists$ and $\forall$.
Instead, we will be talking about a player and its opponent.

\paragraph{Notation.} We write $Q$ for either of the players, and, $\bar Q$ for its opponent.
\vspace{2pt}

Given a QBF $\Q_1X_1,\dots,\Q_nX_n.\,\phi$
the \emph{domain} $\dom(x)$ of a variable $x\in X_k$ are all the variables in the preceding blocks,
i.e.\ $\dom(x)=\bigcup_{i\in 1..k-1} X_i$.

A \emph{play} is a sequence of
assignments $\tau_1,\dots,\tau_n$  where $\tau_i$ is an assignment to~$X_i$.

\begin{definition}\label{def:strategy}
    For a QBF $\Q_1X_1,\dots,\Q_nX_n.\,\phi$ a \emph{strategy for a variable} $x\in X_k$
    is a boolean function $s_x$  whose arguments are the variable's domain, i.e.~$\dom(x)$.

    A \emph{strategy for a player} $\Q$ is a set of strategies $s_x$
    for each of the variables $x\in QX_i$.  Whenever clear from the context, we
    simply say strategy for either of the concepts.
\end{definition}

\paragraph{Notation.}
For the sake of succinctness, a strategy for a variable~$x$ is conflated with a Boolean
formula whose truth value represents the value of the strategy. In another words, a
strategy  represents both some function~$s_x\,:\,2^{\dom(x)}\mapsto\{0,1\}$ and some formula $\psi_x$
with $\vars(\psi_x)\subseteq\dom(x)$.
This convention lets us also treat a set of strategies~$S$ for some variables $X$ as
a substitution.  Hence, $\xi[S]$ represents the formula that results from
simultaneously replacing in $\xi$ each variable $x\in X$ with its strategy~$\psi_x$.

\begin{definition}[winning strategy]\label{def:winningStrategy}
    Let $\Psi$ be a closed QBF $(QX\dots\,\phi)$ with $\phi$ propositional.
    A strategy $S$ for $\exists $ is \emph{winning in} $\Psi$ if $\phi[S]$ is a tautology.
    A strategy $S$ for $\forall $ is \emph{winning in} $\Psi$ if $\phi[S]$ is unsatisfiable.

    In particular, for a formula~$\exists X.\,\phi$ a winning strategy for $\exists$
    corresponds to a satisfying assignment of~$\phi$.
\end{definition}

\begin{observation}\label{obs:winValidity}
    A closed QBF $\Phi$ is true iff there exists a winning strategy for~$\exists$;
    it is false iff there exists a winning strategy for $\forall$.
\end{observation}

\begin{definition}[winning/counter move]\label{def:winningMove}
    For a closed QBF $QX.\,\Phi$ an assignment $\tau$ to $X$ is a \emph{winning move}
    if there exists a winning strategy for $Q$ in $\Phi[\tau]$.

    For a closed QBF $QX\bar QY.\,\Phi$  and an assignment $\tau$ to $X$,
    an assignment $\mu$ to~$Y$ is a \emph{counter-move} to $\tau$
    if $\mu$ is a winning move for $\bar QY.\,\Phi[\tau]$.
\end{definition}

\begin{observation}\label{obs:winMove}
    There exists some winning move for $QX$ in a formula $QX.\,\Phi$,
    if and only if there exists a winning strategy for $Q$ in the formula.
\end{observation}

\begin{observation}
    For a formula $QX\bar QY.\,\Phi$,
    an assignment to $X$ is a winning move if and only if there does not exist a counter-move to it.
\end{observation}

\section{Algorithm \qfun}\label{sec:algo}
As to make it a more pleasant read, this section comes in three  installations, each bringing in more detail.
The first part quickly overviews the existing algorithm (R)AReQS and sketches the main ideas of the proposed approach,   which we will simply call the algorithm \qfun.
The second part presents \qfun for the two-level case, i.e., formulas with one quantifier alternation.
Finally, the third part details out the algorithm for the general case, i.e., formulas with arbitrary number of quantifier alternations.

\subsection{Exposition}

Let us quickly review the existing algorithm \rareqs~\cite{JanotaKMC16}.
For a formula $QX\bar QY.\,\Phi$ \rareqs aims to decide whether  there exists
a winning move for~$Q$.
To that end,  the algorithm keeps on constructing a sequence of pairs
$(\tau_1,\mu_1),\dots,(\tau_k,\mu_k)$.
%
Each $\tau_i$ is an assignment to $X$ and~$\mu_i$ is a counter-move to $\tau_i$ (see Def.~\ref{def:winningMove}).
In each iteration, \rareqs constructs a partial  expansion (called \emph{abstraction}) of the original QBF such that
no existing $\mu_i$ is a counter-move in the original formula to any winning move of the abstraction.
In another words, if $Q$ draws the next move so that it wins the  abstraction,
he is guaranteed not to be beaten by any of the existing counter-moves.

If there is no winning move for the abstraction, there isn't one for the
original formula either and therefore there is no winning strategy for $Q$ (we are done).
If there is some winning move $\tau_{k+1}$  for the  abstraction, we check whether the opponent still comes up with a counter-move $\mu_{k+1}$. If he does not, $\tau_{k+1}$ is a winning move for $Q$ and we are again done (see  Observation~\ref{obs:winMove}).
If a counter-move is found, the pair $(\tau_{k+1},\mu_{k+1})$ is added to the sequence and the process repeats.

This setup inspires the use of machine learning. Since  each $\mu_i$ is a
counter-move to $\tau_i$,  the  constructed sequence of pairs $(\tau_i,\mu_i)$ can be conceived as a \emph{training set} for
the strategies for the variables~$Y$ (belonging to the player $\bar Q$).
More specifically, for  each variable $y\in Y$,
the pair $(\tau_i$,$\mu_i)$ represents a  training sample for the function $s_y$
prescribing that $s_y(\tau_i)=\mu_i(y)$.
Observe that there might be other good strategies for the opponent $\bar Q$. However, the pairs
$(\tau_i,\mu_i) $ have  already \emph{proven} to be good for $\bar Q$ and therefore we will stick to them.

It is tempting to learn a strategy for $\bar QY$ from such samples and then
verify that it is a winning one.  If it is a winning one, we would be done. If
it is not a winning one, we could just learn a better one once we have more
samples.  However, this approach is unlikely to work.
%
%
The problem with this
approach is twofold.  Firstly, it is overly optimistic to hope to hit the right
strategy given a set of samples whose number is likely to be much
smaller than the full truth table of the strategy. Secondly, it is putting
too much strain on machine learning because the set of samples keeps on
growing.  Instead, this paper proposes the following schema.

\begin{enumerate}
    \item Collect some suitable set of samples $\samples$.
    \item Learn strategies $S$ for the opponent variables.
    \item Strengthen the current  abstraction using the strategies $S$.
    \item Reset the set of samples $\samples$
    \item Repeat.
\end{enumerate}

\subsection{\qfunT: 2-level QBF}

Let us look at the two-level case, i.e., a QBF of the form $QX\bar QY.\,\phi$
with $\phi$ propositional. This form is particularly amenable to analysis
since both the abstraction and candidate-checking  is solvable by a SAT solver.
Also, 2-level QBF has a number of interesting applications (cf.~\cite{BenedettiJSAT08,RintanenAAAI07,Umans-sigact02}).

A slight generalization of a game called a
\emph{multi-game}~\cite{JanotaKMC16} is useful in the
following presentation. A multi-game is a set of sub-games where
the top-level player must find a move that is winning for all these sub-games at once.
Note that a multi-game can be converted to a standard QBF by prenexing.
However, it is useful to maintain this form (see~\cite[Sec.~4.1]{JanotaKMC16}).

\begin{definition}[multi-game]
    A multi-game is written as $QX.\,\{\Phi_1,\dots,\Phi_k\}$.
    An assignment $\tau$ to $X$  is a winning move for it
    iff  it is a winning move for all $QX.\,\Phi_i$.
    Each $\Phi_i $ is called a \emph{sub-game} and  is either propositional or  begins with~$\bar Q$.
\end{definition}

\begin{algorithm}[t]
    \nonl\Func \winsOne($QX.\ \left\{\phi_1,\dots,\phi_n\right\}$) \;
    \Input{All $\phi_i$ propositional.}
    \Output{a winning move for all $QX.\,\phi_i$, if there is one; $\nul$ otherwise.}
    $\alpha\gets (Q=\exists)\,?\,\bigwedge_{i\in 1..n} \phi_i\,:\,\bigwedge_{i\in 1..n} \lnot\phi_i$\;
    \Return $\SAT(\alpha)$\;
    \caption{Playing 1-move multi-game}\label{algorithm:w1}
\end{algorithm}

\begin{algorithm}[ht]
    \nonl\Func $\qfunT(QX\bar QY.\,\phi)$ \;
    \Input{$\phi$ is propositional.}
    \Output{a winning move for $QX$ if there exists one, $\nul$ otherwise}
    \BlankLine
    $\samples\gets\emptyset$\tcp*[r]{start with no samples}
    $\alpha\gets\emptyset$\tcp*[r]{empty abstraction}
    \While{\strue}{
        $\tau\gets\winsOne(QX.\,\alpha)$\label{step:candidate2}\tcp*[r]{candidate}
        \lIf(\tcp*[f]{loss}){$\tau=\nul$}{\Return $\nul$}
        $\mu\gets\winsOne(\bar QY.\, \{\phi[\tau]\})$\label{step:counterexample2}\tcp*[r]{countermove}
        \lIf(\tcp*[f]{win}){$\mu=\nul$}{\Return $\tau$}
        $\samples\gets\samples\union\{(\tau,\mu)\}$\tcp*[r]{record sample}
        \If{$\magic()$}{
            $S\gets\learn(\samples)$\tcp*[r]{learn}
            $\alpha\gets\alpha\cup\{\phi[S]\}$\;
            $\samples\gets\emptyset$\tcp*[r]{reset samples}
        } \Else {
            $\alpha\gets\alpha\cup\{\phi[\mu]\}$\tcp*[r]{refine}
        }
    }
\caption{\qfunT: 2-level QBF Refinement with Learning}\label{algorithm:QFUN2}
\end{algorithm}

When all sub-games are propositional, the multi-game is solvable by a single SAT  call.
For such we introduce a function $\winsOne$ (Algorithm~\ref{algorithm:w1}).
The function calculates a winning move for the multi-game or returns $\nul$ if it does not exist
(the function~$\SAT$ has the same behavior).
Observe that if the set of sub-games is \emph{empty}, the
formula $\alpha$ in \winsOne is the empty conjunction, which is equivalent to true, i.e.,
the SAT call then returns an arbitrary assignment.

Just as  the existing algorithm \areqs, \qfunT (Algorithm~\ref{algorithm:QFUN2})
maintains an \emph{abstraction} $\alpha$.
The abstraction corresponds to a partial expansion of the inner quantifier.
This means that for a formula $QX\bar QY.\phi$, the abstraction has the form
$QX.\ \comprehension{\Phi[S]}{S\in\omega}$, where $\omega$ is some set of strategies.
Observe that the abstraction is trivially equivalent to the original formula
if $\omega$ contains all possible constant functions.

For instance, $\forall u\exists e.\,\phi$ is equivalent
to $\forall u.\,\phi[\assgn{0}{e}]\lor\phi[\assgn{1}{e}]$,
which is equivalent to the multi-game
$\forall u.\,\{\phi[\assgn{0}{e}], \phi[\assgn{1}{e}]\}$.

\begin{example}
     Consider $\forall uw\exists xy.\,\phi$
     with $\phi=
     (u\Rightarrow (\lnot w\Leftrightarrow x\land w\Leftrightarrow y))
     \land
     (\lnot u\Rightarrow (w\Leftrightarrow x\land \lnot w\Leftrightarrow y))
     $.
     The following  abstractions of this formula are both losing for $\forall$.
     With two sub-games:
     $\forall uw.\,
     \{ \phi[\assgn{\lnot w}{x}, \assgn{w}{y}],\phi[\assgn{w}{x}, \assgn{\lnot w}{y}] \} $;
     with single sub-game:
     $\forall uw.\,
     \{ \phi[\assgn{(\ite{u}{\lnot w}{w})}{x},\assgn{(\ite{u}{w}{\lnot w})}{y}] \}$.
\end{example}

The abstraction $\alpha$ is \emph{refined} with every play losing for~$Q$,
which effectively means adding a subgame to the current abstraction.
Additionally, \qfunT maintains a set of samples~$\samples$.
The samples are pairs $(\tau_i,\mu_i)$ such that $\tau_i,\mu_i$ is a losing play for $Q$,
i.e., a winning play for $\bar Q$.
So for instance, if $Q=\forall$ then $\bar Q=\exists$ and $\tau_i\union\mu_i\models\phi$.

Both the abstraction~$\alpha$ and samples~$\samples$ are initialized as empty. In each iteration,
\qfunT calls \winsOne to calculate a \emph{candidate} for a winning move~$\tau$.
Subsequently, another call to~$\winsOne$ is issued to  calculate a counter-move~$\mu$.
If either candidate or counter-move does not exist, one of the player has lost without recovery.

Machine learning is invoked only ever so often. To decide when, the
pseudo-code  queries the function $\magic$.
Whenever $\magic$ is true, new
strategies are learned for $Y$-variables based on the samples $\samples$.
These strategies are plugged into the formula~$\phi$ and recorded in the
abstraction.  Then, and that is crucial, the set of samples is reset back to
the empty set.

In terms of  soundness, the set  of samples $\samples$ need not be reset after each learning.
However, it is crucial in terms of performance.
If the set is always augmented, learning will become overly time consuming.
Recall that a strategy needs to be learned for each opponents' variable and further, the number of iterations can go to millions.

So what does the abstraction represent and what is the role of the learned
strategies?  The original \areqs adds a sub-game $\phi[\mu_i]$ for each
existing counter-move $\mu_i$.
Intuitively, this means that the player~$Q$ never plays before making sure that
he can successfully defend himself against all the existing counter-moves. Once
strategies are also included, the player~$Q$ also defends himself against all the
strategies devised so far.  Strategy-based refinement is a
generalization of the traditional one---the traditional refinement corresponds
to a set of strategies comprising constant functions.

What do we require from the strategy learning? The good news is that in fact
very little.
A strategy must be learned in the form of a formula so that it can be plugged
into the original formula. This means that the algorithm does not easily allow for
neural networks, for instance. The current implementation uses decision-trees (see
Section~\ref{sec:implementation}).
For the sake of soundness, the learned  strategy formula \emph{must follow} the definition of a strategy (Definition~\ref{def:strategy}).
In practice this means that a strategy formula $\xi_y$
for a variable $y$ must only contain variables from $\dom(y)$.

If the function $\magic$   triggers traditional refinement only finitely many
times, the learning method also needs to guarantee termination of the  whole algorithm.
A natural minimal requirement  for this is that the learned
strategies will correspond to  at least one sample $(\tau_i,\mu_i)\in\samples$,
i.e.\ $s_y(\tau_i)=\mu_i(y)$, for each $y\in Y$. This requirement guarantees that
$\tau_i$ will not appear as a candidate  for a winning move in the upcoming
iterations. Nevertheless, if $\magic$ alternates between traditional and
learning-based refinement, termination is already guaranteed by the traditional
refinement and we do not need to worry about what is learned as long as it is sound.

\subsection{\qfun: General Case}
\begin{algorithm}[t]
    \nonl\Func $\qfun(QX.\,\left\{\Phi_1,\dots,\Phi_n\right\})$ \;
    \Input{Each $\Phi_i$ is propositional or begins with $\bar QY$.}
    \Output{a winning move for $QX$ if there exists one, $\nul$ otherwise}
    \BlankLine
    \If{all $\Phi_i$ propositional}{\Return$\winsOne(QX.\,\left\{\Phi_1,\dots,\Phi_n\right\})$}
    $\samples_i\gets\emptyset$, $i\in 1..n$\tcp*[r]{samples}
    $\alpha\gets QX.\emptyset$\tcp*[r]{empty abstraction}
    \While{\strue} {
        $\tau'\gets\qfun(\alpha)$\label{step:candidate}\tcp*[r]{candidate}
        \lIf(\tcp*[f]{loss}){$\tau'=\nul$}{\Return $\nul$}
        $\tau\gets\comprehension{l}{l\in\tau' \land \var(l)\in X}$\label{step:multi_filter}\tcp*[r]{filter}
        \lIf(\tcp*[f]{win}){all $\qfun(\Phi_i[\tau])=\nul$}{\Return$\tau$}
        \Let $l$ be s.t.\ $\qfun(\Phi_l[\tau])=\mu$ for some $l\in\{1..n\}$, $\mu\neq\nul$\;
        $\samples_l\gets\samples_l\union\{(\tau,\mu)\}$\tcp*[r]{record sample}
        \If{$\magic()$}{
            $S\gets\learn(\samples_l)$\tcp*[r]{learn}
            $\alpha\gets\refine(\alpha,\Phi_l, S)$\;
            $\samples_l\gets\emptyset$\tcp*[r]{reset samples}
         } \Else {
             $\alpha\gets\refine(\alpha,\Phi_l, \mu)$\tcp*[r]{refine}
         }
    }
\caption{QBF Refinement with Learning}\label{algo:general}
\end{algorithm}

The general case \qfun generalizes  the two-level case \qfunT using recursion (just as \rareqs generalizes \areqs).
The basic ideas remain,  even though we are faced with a couple of technical complications.
The pseudocode is presented as Algorithm~\ref{algo:general}.
Since the abstraction is a multi-game, the recursive call also needs to handle a multi-game.
For this purpose, we maintain a set of sequences of samples---each sequence for each given sub-game.
Candidates for a winning-move are drawn from the abstraction $\alpha$ by a recursive call.
The small technical difficulty here is that the abstraction may return a winning move containing some extra
fresh variables coming from refinement. Hence, these need to be filtered  out (ln.~\ref{step:multi_filter}).

If, the candidate  move $\tau$ is a winning move, it is returned.
If, however, there is some  counter-move $\mu_i$ obtained by playing the sub-game $\Phi_i$, it is used for refinement.
This means inserting the pair $(\tau,\mu_i)$ into the sample sequence pertaining to this sub-game, i.e.\ sequence $\samples_i$.
And subsequently, performing refinement.  In order to ensure that quantifiers
alternate, refinement  introduces fresh variables for formulas with more than~2
levels. The refinement function is defined as follows.

\noindent
\begin{minipage}{.45\textwidth}
$\refine \big(Q X.\{\Psi_1,\dots,\Psi_n\},\ \bar Q Y Q X_1.\,\Psi,\ S\big) \ :=\   Q X X_1'.\{\Psi_1,\dots,\Psi_n,  \Psi'[S]\}$
\end{minipage}

\noindent
\begin{minipage}{.45\textwidth}
$\refine \big(Q X.\{\Psi_1,\dots,\Psi_n\},\ \bar Q Y.\,\psi,\ S\big) \ :=\  Q X.\{\Psi_1,\dots,\Psi_n, \psi[S]\}$
\end{minipage}

\noindent\hspace{3pt}
\begin{minipage}{.45\textwidth}
\it where $X_1'$ are fresh duplicates of the variables~$X_1$ and~$\Psi'$ is~$\Psi$  with~$X_1$  replaced by~$X_1'$
and where $\psi$  is a propositional formula.
\end{minipage}

\section{Implementation}\label{sec:implementation}

\subsection{Formula representation}
The algorithm requires nontrivial formula manipulation to achieve refinement.
Performing these operations directly on a CNF representation is difficult
and further, CNF representation as input has well-known pitfalls~\cite{selman-aaai05}.
Hence, the implementation  represents formulas as \emph{And-Inverter
graphs}~(AIG)~\cite{Hellerman63}, which are simplified by trivial
non-invasive simplifications~\cite{BrummayerBiere06}.
%
All the logical operations (e.g.\ substitution/conjunction) are performed on
AIGs.  Only when the time comes to call a SAT solver, the AIG is translated
into CNF.  This is done in straightforward fashion. Each sub-AIG is
mapped to an encoding Boolean variable in the SAT solver.  Since the AIGs are
hash-consed, each sub-AIG also corresponds to just one variable.  All the
and-gates  are binary.  The input to the solver is the circuit-like format for
QBF called \emph{QCIR}~\cite{klieberBP16}.

\subsection{Learning}
Recall that learning is invoked with the sequence of pairs of assignments
$\samples = (\tau_1,\mu_1),\dots,(\tau_k,\mu_k)$,
where each $\tau_i$ is an assignment to some block of variables $X$ in the prefix and
$\mu_i$ is an assignment to variables $Y$, which is the adjacent block in the prefix,  belonging to the opposing player.

The objective is to learn a strategy (a function) for each of the variables in $Y$.
A Boolean function can be seen as a \emph{classifier} with two classes:
the input assignments where the strategy should return 1 (true)
and the input assignments where the strategy should return 0 (false).
The implementation uses the popular classifier \emph{Decision trees}~\cite{RussellNorvig10}.
These are constructed by the standard \emph{ID3} algorithm~\cite{quinlan86}.

For each variable in $y\in Y$, construct the training set $\samples_y$ from $\samples$  by ignoring all the other $Y$ variables.
Subsequently invoke ID3 on $\samples_y$ thus obtaining a decision tree conforming to the  sample assignments.
Once a decision-tree is constructed, the  Boolean formula is constructed as follows.
\begin{enumerate}
    \item Construct the sets of conjunctions of literals $\Is_p$ and $\Is_n$
        corresponding to the positive and negative branches of the tree, respectively.
        Hence, if~$t\in\Is_p$ is true, the tree gives~$1$.
    \item\label{it:sub} Repeatedly apply subsumption and self-subsumption on each set $\Is_p$ and $\Is_n$, until a fixed point is reached.
    \item\label{it:smaller}  If $|\Is_p|<|\Is_n|$ return $\bigvee\Is_p$, otherwise return $\lnot\bigvee\Is_n$.
\end{enumerate}

Step~\ref{it:sub} would not necessarily be needed but since we are substituting the constructed functions into the input formula,
it is desirable to maintain them small. Analogously, either set could be chosen in step~\ref{it:smaller} but a smaller is preferable.

\paragraph{When to learn?}
 It is a bad idea to learn too frequently since this would produce poor sample-sets to learn from.
 However, learning too \emph{infrequently} has two main pitfalls:
 \begin{enumerate}
     \item  Learning on large sample-sets will be too costly (recall that a learning algorithm is run for each opponent variable upon refinement).
     \item There is a risk of very complicated and therefore large functions to be learned from complicated samples
 \end{enumerate}
A straightforward  approach was taken to implement the function $\magic$: learning is triggered every $K$ iterations of the loop,
where $K$ is a parameter of the solver. The number of iterations is considered local for each recursive call of $\qfun$.
The experimental evaluation examines the solver's behavior for several values of~$K$ (see Section~\ref{sec:experiments}).

\subsection{Strategy accumulation}\label{sec:accumulation}
Upon each refinement the set of samples is reset. Also, whatever is learned is
forgotten in the next rounds---learning starts from scratch on a new set of
samples. This might be disadvantageous. The current implementation uses a
simple but important improvement.   The algorithm records for each variable~$y$ the last learned strategy.
This strategy is then evaluated on the next batch of samples
when learning is invoked again.  If it still fits the data, it is kept. Otherwise it is discarded and a new  strategy is learned.

\begin{example}\it
     Consider the formula from the introduction of the paper:\\
     $\forall x_1,\dots,x_n\exists y_1,\dots, y_n.\,\bigwedge x_i\leftrightarrow y_i$,
     and, the following sequence of samples.

      \rm
      \noindent\begin{minipage}{.46\textwidth}
          \centering
          \begin{tabular}{|c|c|c|c||c|c|c|c|}
              \hline
              $x_1$ & $x_2$ & $\dots$ & $x_n$ & $y_1$& $y_2$ & $\dots$ & $y_n$\\\hline\hline
              0     &   0   & $\dots$ & 0   & 0  & 0 & $\dots$ & 0\\\hline
              1     &   0   & $\dots$ & 0   & 1  & 0 & $\dots$ & 0\\\hline\hline
              0     &   0   & $\dots$ & 1   & 0  & 0 & $\dots$ & 1\\\hline
              0     &   1   & $\dots$ & 1   & 0  & 1 & $\dots$ & 1
          \end{tabular}
      \end{minipage}
      \it

      If $K=2$, the first application of  learning gives $y_1\triangleq x_1$ and the rest of the  strategies are constantly 0.
      In the second refinement, learning gives $y_2\triangleq x_2$ and the rest constants.
      If, however,  we keep the information from the previous
      learning, we get both  $y_1\triangleq x_1$, $y_2\triangleq x_2$.
      Hence, accumulating the individual strategies will eventually  yield the right strategy.
\end{example}

\subsection{Incrementality}
The recursive structure of the algorithm is very elegant but might be too
forgetful.  If one is to solve $\Phi[\mu_i]$,  it could be useful to maintain
the  abstraction from that solving in order to solve $\Phi[\mu_{i+1}]$. The
issue is that then the solvers tend to occupy too much space. Currently, the solver
maintains only abstractions that are purely propositional.

\section{Experimental Evaluation}\label{sec:experiments}
\begin{table}[t]
    \centering
    \begin{tiny}
    \begin{tabular}{|l|c|c|c|c|c|c|c|c|}
        \hline
        \textbf{Solver} & \textsf{Quabs} &   \textsf{GQ} & \textsf{RAReQS} & \textsf{L-16} &  \textsf{L-64} & \textsf{L-128} & \textsf{L-64-f}\\\hline
        \textbf{Solved (320)} & 103 & 75    & 105    & 110 & \textbf{111} & \textbf{111} & 104 \\\hline
        \textbf{Wins}   & 63  &  11    &  \textbf{67}   &  55 &  63 &  62 & 60 \\\hline
    \end{tabular}
    \end{tiny}
    \caption{Result summary. A win is counted also for solvers that are not worse than the best time by 1s.}\label{tab:results}
\end{table}

\begin{figure}[t]
    \centering
        \include{qbfeval17}

    \begin{minipage}{.4\textwidth}%
        {\small\textbf{Fig.\,1a} Cactus plot. A point at $(x,y)$ means that the solver solved $x$ instances each within $y$~sec.}
    \end{minipage}


        \include{64VSd-its}
        {\small\textbf{Fig.\,1b} Log-scale scatter plot for \#iterations.}
\end{figure}

The \rareqs algorithm has proven to be highly competitive
as it have placed first in several tracks of the recent QBF competitions.\footnote{\url{http://www.qbflib.org/}}
So the key question is whether \rareqs benefits, or may benefit, from  the  proposed learning.

The success of the machine learning techniques can be assessed at various levels.
The lowest bar is whether  the technique is at all \emph{computationally feasible}.
Indeed, it  might be that the learning  is impractically time-consuming.
Second step is whether the \emph{number of iterations decreases when learning is applied}.
The third step is whether also \emph{solving time decreases when learning is applied}.
Finally, we are interested in variations of the algorithm. Namely, the effect of the learning interval
and the effect of the technique of accumulating strategies (see Section~\ref{sec:accumulation}).


The evaluation considers the following  configurations of  algorithm \qfun (Algorithm~\ref{algo:general}):
\qfun \emph{without} any learning, which is in the fact \rareqs;
versions \qfun-16, \qfun-64, and \qfun-128 where learning is triggered every 16/64/128 iterations, respectively;
\hbox{\qfun-64-f} \textbf{f}orgetful version of \qfun where previously learned strategies are not used in the future.
All the other versions accumulate strategies as described in Section~\ref{sec:accumulation}.

Additionally we compare to the highly competitive non-CNF solvers
\textsf{GhostQ}~\cite{KlieberSAT10} and \textsf{QuAbS}~\cite{TentrupSAT16}.

The prototype used for the evaluation is implemented in $C^{++}$ and the SAT
solver \textsf{minisat 2.2}~\cite{EenS03} is used as the backend solver.  The
experiments were carried out on Linux machines with Intel Xeon 5160 3GHz
processors and 4GB of memory with the time limit $800$\,s and memory limit~2GB.

For the evaluation we used the non-CNF suite from the \emph{2017 QBF Competition} counting 320 instances.
\footnote{\url{http://www.qbflib.org/event_page.php?year=2017}}

The  overall results are summarized in Table~\ref{tab:results}.
The cactus plot in Fig.~1a summarizes the performance.
For the sake of readability, the cactus plots omits \qfun-16,
 whose performance  is quite similar to \qfun-64 and \qfun-128, which are already quite close.
Fig.~2b is a scatterplot comparing the total number of refinements for \textsf{\qfun-64} and \textsf{\rareqs},
i.e., machine learning every 64 iterations versus no learning.
Detailed results are provided as supplementary material.

\subsection{Results Discussion}

Overall, learning  gives improvement both in terms of  number of solved instances as well as number of iterations.
Admittedly, in terms of number of solved instances the gain is modest.
However, the difference in performance between RAReQS and QuAbS is even smaller despite
each representing a completely different algorithm.
Also recall that Fig.~2b is in logarithmic scale so the number of iterations saved are in number of cases in orders of magnitude.
Overall this suggests that adding learning in the brings about a new quality in the solver.

The effect of frequency of learning on the performance is relatively small.
The  best configuration is with learning every 64 refinements (\qfun-64),
while \qfun-16 and \qfun-128 perform slightly worse.
This is not surprising as too frequent learning will slow down the solving and too infrequent
does not give enough opportunity to learn.

The biggest effect has strategy accumulation. Indeed, without it, learning in
fact performs worse then without any learning.  This suggests that at least for
some variables it is important to learn a certain strategy and maintain it.
This observation clearly opens opportunities for further investigation as the
techniques of accumulating strategies can be further developed.

\section{Related Work}\label{sec:related}
The research on QBF solving has been quite active in the last decades and
an array of approaches exists.  It appears that these different approaches also
give us a different classes of instances where they are successful.  One of the oldest
approaches is conflict/solution learning~\cite{zhang-iccad02,lonsing-jsat10,giunchiglia2010system,lonsing-thesis12}, which essentially generalizes clause learning in SAT.
Then there are solvers that perform quantifier expansion into Boolean
connectives~\cite{Biere04,Benedetti05,LonsingNenofex08,PigorschScholl-dac10,TuHJ15}; solvers that
target non-CNF
inputs~\cite{zhang-aaai06,KlieberSAT10,GoultiaevaB10,Goultiaeva-date13,Gelder-cp13,BalabanovSAT16,TentrupSAT16};
and solvers that  calculate blocking clauses using a SAT
solver~\cite{malik2qbf,JanotaIJCAI15,RabeFMCAD15}.
Recently we have also seen integration of inprocessing with conflict/solution learning~\cite{LonsingSAT16}.

This paper builds  on the algorithm \rareqs~\cite{JanotaKMC16},
which expands quantifiers gradually by substituting them one by one into the formula.
This approach is conceptually akin to the \emph{model-based quantifier instantiation}~\cite{WintersteigerFMSD13}.

It is known that QBF solvers \emph{implicitly} trace  strategies because a winning
strategy can be extracted once the formula is
solved~\cite{Goultiaeva-ijcai11,balabanov-fmcad12,BalabanovJJW-aaa15,BCJ14}.
However, to our best knowledge there are  currently only two QBF solvers that \emph{explicitly}
target strategy computation.  In~\cite{janota-lpar15} the authors
fused clause learning and \rareqs by refining abstractions with strategies
calculated from clause learning---with not very promising results.
The second solver by Rabe and Seshia works in the context of 2QBF and
gradually adds variables to a winning strategy of the inner quantifier~\cite{RabeSAT16}.

It is hard to do justice to the work that has been done in machine
learning, the reader is directed to standard literature~\cite{RussellNorvig10}.
It should be mentioned that strategy learning is a very specific type of learning because we need the result in the form of a formula. This is closely related to function synthesis/learning cf.~\cite{Valiant84,Resende92,OliveiraNIPS93,SuMLSP16}.
Machine learning has also been used in portfolio solvers e.g.~\cite{hoos-jair08}
or to predict formulas' value~\cite{hoos-aaai12}.

Last but not least, machine learning has been used at a higher level of
inference to discover lemmas in the context of first order or higher order
reasoning~\cite{UrbanIJCAR08,urbanFlyspeck}.


\section{Conclusion and Future Work}\label{sec:conclusion}

This paper presents a QBF solver that  periodically \emph{generalizes} a set of observations (plays)
into a strategy by machine learning. These strategies are plugged into the
original formula in order to gradually strengthen  a partial expansion of the formula.
The results show that this is feasible and it also helps to reduce
the number of refinement iterations but also the solving time.
The fact that this results in a competitive QBF solver is already  compelling.
Indeed, machine learning is invoked many times during solving on a number of variables separately.
However, the design of the algorithm enables us to curb the computational burden of machine learning by
limiting the size of the training set.

As discussed in Section~\ref{sec:implementation}, the current prototype
is rather straightforward in its  implementation decisions.  There is a lot of
room  for making the   solver more intelligent.  Besides inprocessing and
other implementation issues, number of things are to be investigated for the
machine learning part.  What kind of machine learning methods are good for this
purpose?  When to trigger machine learning?  Can we improve the training sets (e.g.\ introduction of don't-cares)?

Another interesting question for future work is whether machine learning can
be beneficial in other type of QBF solving.  There are opportunities for
this. Even if the solver is \emph{not} performing expansion-based refinement
(e.g.\ CAQE~\cite{RabeFMCAD15}, QESTO~\cite{JanotaIJCAI15}, CADET~\cite{RabeSAT16}),
it can for instance use a learned strategy to predict the behavior of the opponent.

At the theoretical level, the paper touches a fundamental question:
how difficult is it to learn the right strategies? Here, PAC-learnability
could give some answers~\cite{Valiant84}.

\bibliographystyle{splncs03}
\bibliography{refs,learning}
\end{document}